# Electrical and Hydrogen Reduction Enhances Kinetics in Doped Zirconia and Ceria: II. Mapping Electrode Polarization and Vacancy Condensation in YSZ

Yanhao Dong and I-Wei Chen[*]

*Department of Materials Science and Engineering, University of Pennsylvania,*

*Philadelphia, PA 19104, USA*

**Abstract**

Knowing the correlation between grain boundary mobility and oxygen potential in yttria stabilized zirconia (YSZ), we have utilized the grain size as a microstructural marker to map local oxygen potential. Abrupt oxygen potential transition is established under a large current density and in thicker samples. Cathodically depressed oxygen potential can be easily triggered by poor electrode kinetics or in an oxygen-lean environment. Widespread cavitation in the presence of highly reducing oxygen potential suggests oxygen vacancy condensation instead of oxygen bubble formation as commonly assumed for solid oxide fuel/electrolysis cells. These results also suggest electrode kinetics has a direct influence on the microstructure and properties of ceramics sintered under a large electric current.

[*]Corresponding Author
*E-mail address*: iweichen@seas.upenn.edu (I-Wei Chen)

## I. Introduction

In the electrochemical literature, electrode polarization refers to a discontinuous change in some potential across the electrode in an electrochemical cell when the electrode is under a current.[1,2] (This definition is not to be confused with dipolar polarization in dielectrics and ferroelectrics, which is a vector instead of a scalar, as for electrode polarization above.) The existence of polarization is believed to be quite common in devices such as solid oxide fuel cell[3,4], electrolysis cell[5], and batteries, and it arises because of the need for an extraordinarily large driving force to motive sluggish electrode reactions. In kinetics, this situation is known as the limit of interface control, *vis-a-vis* the other limit, of diffusion control. It is a well-established notion in the phenomenological theory of kinetics, and it is often taken for granted without any direct microstructural evidence, which is especially true for electrode polarization in practical electrochemical devices. This is because device materials are typically pre-screened for their stability under pertinent electrochemical conditions; therefore, they tend to suffer little microstructure changes despite large variations in interface kinetics. Against this background, we aim to provide direct microstructural evidence for electrode polarization in a well-known solid-oxide electrolyte, 8 mol% yttria stabilized zirconia (abbreviated as 8YSZ). Specifically, here the electrode polarization refers to the variation in the local oxygen potential in a zirconia cell.

Our experimental design seeks to take advantage of a key observation in **Part I** of this series of papers[6]: Faster grain growth is correlated to a depressed oxygen potential in 8YSZ and related fluorite-structure ceramics[6-9]. This is schematically shown in **Fig. 1a** for grain boundary mobility (solid line in purple), together with the conductivities of oxygen ions, electrons and holes[10], which are all affected by oxygen potential. The practical significance of this finding is that grain growth kinetics can now be used as a marker of the local oxygen potential. By tracking oxygen potential using grain size, one can determine the polarization distribution inside the electrolyte. In practice, we will simply electrically load YSZ ceramics in different ways and in various atmospheres to create the polarization of interest, then determine their grain size distributions that track their polarizations. Naturally, we expect a microstructure with a graded grain size in electrically loaded YSZ, which has been seen in other studies in our laboratory[6,7].

Our experimental design is specifically guided by the following considerations summarized in **Fig. 1b**. As already mentioned, electrode polarization arises to provide an over potential to compensate for sluggish electrode reactions. Therefore, any drop in the potential—in our case, the oxygen potential $\mu_{O_2}$—should always be on the downstream side of the oxygen flow, which is from the cathode to the anode in **Fig. 1b**. If the YSZ cell is placed inside a known atmosphere to keep the oxygen potential outside the two electrodes, $\mu_{O_2}^{I}$ and $\mu_{O_2}^{II}$, the same, then the potential distribution can be schematically drawn for the four cases shown in **Fig. 1b**. (For more details, see **Appendix**.[11]) (i) Ideal electrodes: Electrode kinetics is so fast that there is no need for over potential, i.e., there is no discontinuity at either electrode. (ii) In air, with adequate electrode kinetics: With "good" electrodes, an "good" electrode catalyst and/or a small current density, electrode reactions can be accommodated by a set of modest over potentials. (iii) In hydrogen: Cathode reactions—to incorporate oxygen from an oxygen-poor environment—are so disfavored that a large over potential is needed; in contrast, anode reactions—releasing oxygen into an oxygen-lean environment—are so favored that there is at most a need for a minimal over potential. (iv) In air, with inadequate electrode kinetics: With either two "bad" electrodes where interface reactions are poorly catalyzed or a very large current density, large over potentials are needed to accelerate the reactions, in order to match the imposed current density. These cases illustrate four possibilities of the potential drop between $\mu_{O_2}^{I}/\mu_{O_2}^{II}$ and its counterpart across the electrode, $\mu_{O_2}'/\mu_{O_2}''$. Although they are not exhaustive, such cases can be readily implemented in the following experiments to examine the interplay between grain growth, current density and atmosphere. It turns out extensive internal reactions between defects in the form of cavitation are also evident in these experiments.

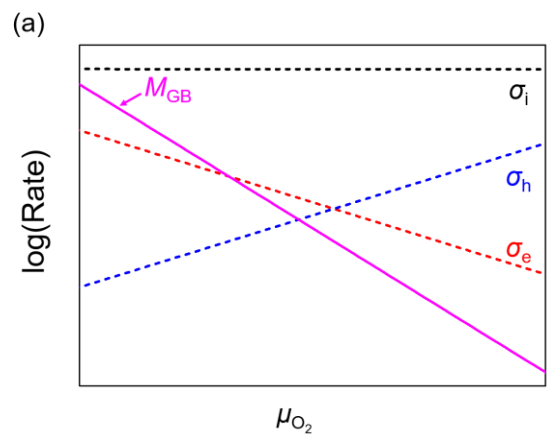

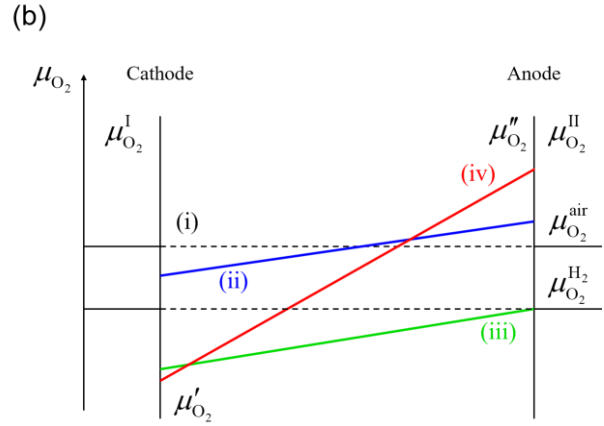

**Figure 1** (a) Schematic dependence of ionic conductivity $\sigma_i$, electron conductivity $\sigma_e$, hole conductivity $\sigma_h$ and grain boundary mobility $M_{GB}$ on oxygen potential $\mu_{O_2}$. (b) Schematic variation of oxygen potentials inside a YSZ cell: (i) ideal electrodes (dash lines in black), (ii) in air with small current density or average electrodes, $\mu'_{O_2} < \mu^{I}_{O_2} = \mu^{air}_{O_2}$ and $\mu''_{O_2} > \mu^{II}_{O_2} = \mu^{air}_{O_2}$ (solid line in blue), (iii) in hydrogen, $\mu'_{O_2} < \mu^{I}_{O_2} = \mu^{H_2}_{O_2}$ and $\mu''_{O_2} \approx \mu^{II}_{O_2} = \mu^{H_2}_{O_2}$ (solid line in green), and (iv) in air with large current density or poor electrodes, $\mu'_{O_2} < \mu^{I}_{O_2} = \mu^{air}_{O_2}$ and $\mu''_{O_2} > \mu^{II}_{O_2} = \mu^{air}_{O_2}$ (solid line in red).

## II. Experimental procedure

Pellets made of cubic $ZrO_2$: 8mol% $Y_2O_3$ (8YSZ) powders (TZ-8Y, Tosoh Co., Tokyo, Japan) were isostatically pressed at 300 MPa at room temperature and sintered at 1300 °C for 12 h to full density with a uniform microstructure and a grain size of 1.7 μm (**Fig. 2a**). They were cut into ~1 mm slices for subsequent grain growth studies under electrical loading with two types of electrodes. One was a uniform coverage of pure Pt paste (Pt-I, Nexceris, LLC., OH) which affixed two Pt wires

(diameter: 0.127 mm) to two opposing sides of the sample; it is the same type of electrode used in our previous studies[6] and was put into use after a heat treatment at 1000 °C for 2 h. The other was made of Pt wires (diameter: 0.127 mm) wrapped around two ends of the sample; it provided limited contact and was used to mimic "bad" electrodes with poor catalytic activities. Electrical loading of the sample, placed in a tube furnace in various atmospheres (air, flowing argon, or flowing forming gas—5% $H_2$+95% $N_2$), was applied by a constant voltage or current from a power supply (G750 potentiostat, GAMRY Instrument, Warminster, PA or Keithley 2400 Source Meter, Keithley Instrument Inc., Cleveland, OH). The *I-V* curve was continuously monitored during the test. Tested samples were cut, polished, thermally etched and examined under a scanning electron microscope (SEM, Quanta 600, FEI Co. Hillsboro, OR) to measure the grain size by the linear intercept method with more than 100 intercepts and a correction factor of 1.56.

**III. Results**

(1) Lowering $\mu_{O_2}^{I}$ and $\mu_{O_2}^{II}$ using reducing atmospheres

Dense samples with an initial grain size of 1.7 μm (**Fig. 2a**) were first annealed at 1300 °C for 4 h to established the reference grain size of 2.1 μm in air (**Fig. 2b**) and 2.7 μm in 5% $H_2$+95% $N_2$ (**Fig. 2c**). To check whether YSZ is well equilibrated with the reducing environment and whether there exists a catalytic effect, we further compared the grain sizes in samples with and without a Pt coating, which is likely to be catalytic, both annealed at 1300 °C for 4 h in 5% $H_2$+95% $N_2$. No difference in grain size was found between them, which established 2.7 μm as the reference grain size under the hydrogen annealing condition. With the above references, we next held the samples with the same initial grain size in 5% $H_2$+95% $N_2$ at the same temperature (1300 °C) and duration (4 h) at different current densities (1, 5 and 10 A/cm$^2$). Different grain sizes were obtained on the cathode side and the anode side, as shown by the solid lines in **Fig. 3**. As also shown there, parallel experiments were performed in argon atmosphere at several current densities (1, 5 and 10 A/cm$^2$, shown by dash lines in **Fig. 3**), which resulted in similar but slightly smaller grain sizes. Selected microstructures of electrically tested samples in 5% $H_2$+95% $N_2$ are shown in **Fig. 4(a-f)**, and the ones tested in argon shown in **Fig. 5(a-f)**.

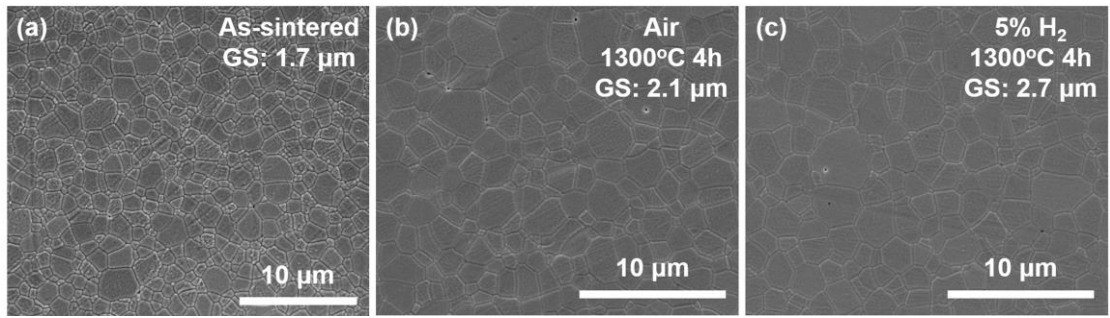

**Figure 2** Microstructure of (a) as-sintered 8YSZ sample, annealed at 1300 °C for 4h in (b) air and (c) flowing 5% $H_2$+95% $N_2$. Grain sizes are listed on the upper-right corners.

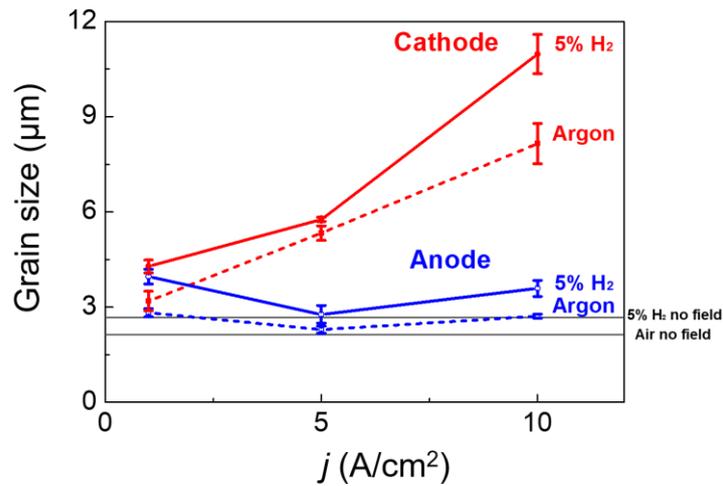

**Figure 3** Grain size on anode (blue) and cathode (red) side of 8YSZ samples electrically tested at 1300 °C in argon and 5% $H_2$+95% $N_2$ under several current densities. Also shown are two reference lines (grey) of samples similarly annealed in air and 5% $H_2$+95% $N_2$ without current. All samples started with same initial microstructure.

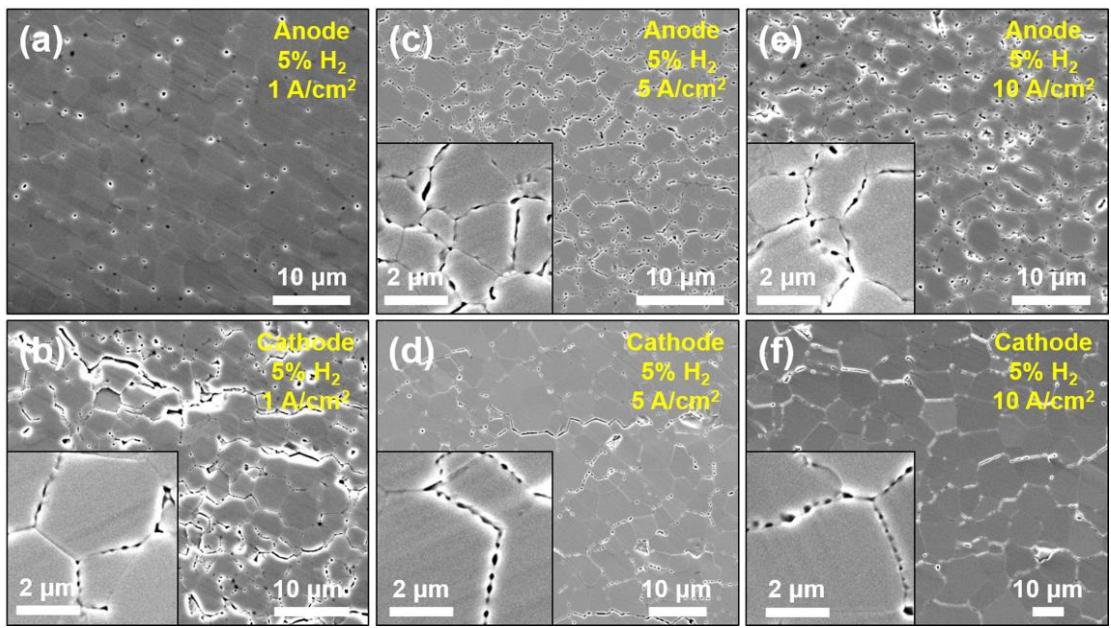

**Figure 4** Microstructures of 8YSZ samples electrically loaded in 5% $H_2$ for 4 h under (a-b) 1 A/cm², (c-d) 5 A/cm², (e-f) 10 A/cm². Furnace temperature: 1300 °C; voltage: (a-b) 1.8 V, (c-d) 3.1 V, (e-f) 4.3 V; sample thickness: (a-b) 0.8 mm, (c-d) 1.3 mm, (e-f)

0.9 mm. Enlarged images with same scale also shown as insets in (b-f).

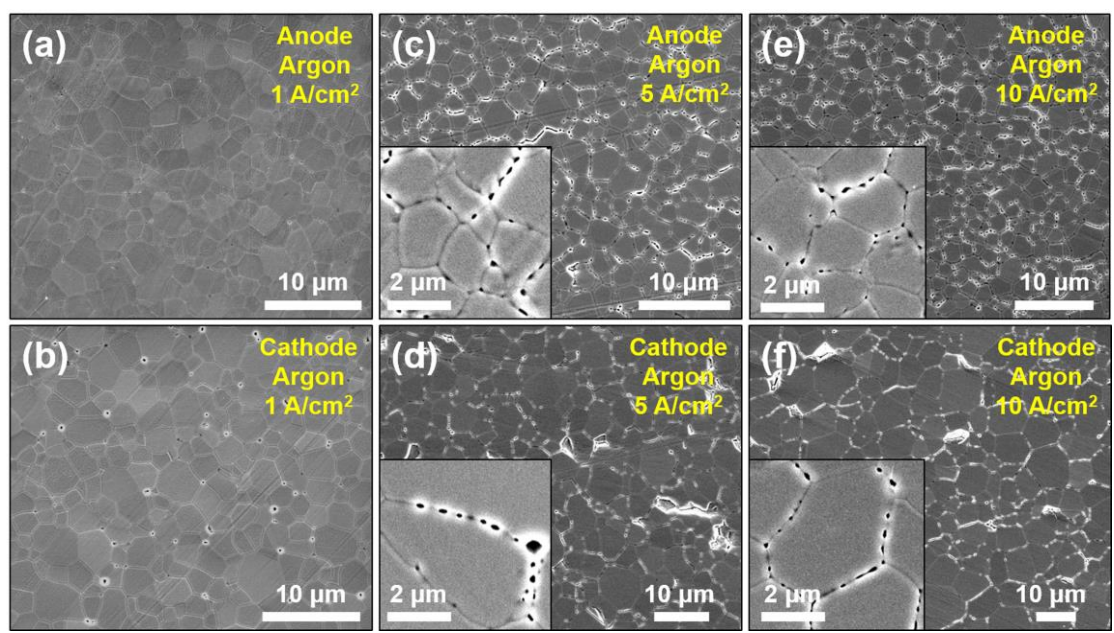

**Figure 5** Microstructures of 8YSZ samples electrically loaded in Ar for 4 h under (a-b) 1 A/cm$^2$, (c-d) 5 A/cm$^2$, (e-f) 10 A/cm$^2$. Furnace temperature: 1300 °C; voltage: (a-b) 2.2 V, (c-d) 3.3 V, (e-f) 4.6 V; sample thickness: (a-b) 1.0 mm, (c-d) 1.1 mm, (e-f) 0.9 mm. Enlarged images with same scale also shown as insets in (c-f).

The measured grain sizes have the following features: (i) In all cases, the grain sizes on the cathode side (in red) are larger than those on the anode side (in blue), which are in turn larger than the reference grain sizes without an electrical load; (ii) the grain size on the cathode side increases rapidly with the current density; (iii) the grain sizes after electrical testing in 5% H$_2$+95% N$_2$ (solid lines) are larger than the ones after testing in argon (dash lines, less reducing). Compared to the results in **Paper I**[6], the onset of grain-growth enhancement occurs at a smaller current density: Previously, there was no apparent grain growth on the cathode side when tested in air unless 20 A/cm$^2$ was exceeded (see **Fig. 7b** of Ref. 7 and **Fig. 11** of Ref. 6), there is now obvious grain growth at less than 10 A/cm$^2$ when tested in 5% H$_2$. Therefore, on the cathode side, the combination of atmospheric reduction and electric current has enhanced grain growth much more than either one alone. This signals that a large over potential has been developed at the cathode when the YSZ cell is electrically loaded in a reducing atmosphere, corresponding to case (iii) in **Fig. 1b**.

Interestingly, on the anode side the grain size initially decreases with current density, reaching a minimum at 5 A/cm$^2$. This turns out to be caused by boundary-pinning porosity, which was developed in all the electrically tested samples, including at 1 A/cm$^2$ (**Fig. 4-5**). Since severe Joule heating can lead to thermal runaway, which makes it difficult to know the bulk temperature[12-14], the 1 A/cm$^2$ data which had experienced minimal Joule heating are especially informative. In both argon and 5% H$_2$+95% N$_2$ annealing, the small electrical load of 1 A/cm$^2$ yielded a larger grain size than the reference (upper grey line in 5% H$_2$+95% N$_2$). This suggests a more reducing condition than the atmosphere has also developed on the anode side, $\mu''_{O_2} < \mu^{II}_{O_2} = \mu^{H_2}_{O_2}$, which is not indicated in the simplified schematic case (iii) in **Fig. 1(b)**.

Lastly, grain size distributions across the electrolyte are shown for selected cases in **Fig. 6**. Here, **Fig. 6a-b** are from our previous studies of high-current air testing[6,7], which develops a sharp grain-size transition in the middle section. This transition is still relatively sharp in **Fig. 6c** when the current density is reduced to ~10 A/cm$^2$, again tested in air and with a poor electrode of Pt wire (see **Fig. 7** to be described in the next subsection). In comparison, at the same 10 A/cm$^2$ the grain size variation in the present study in 5% H$_2$ is rather gradual in **Fig. 5d**, and it becomes even more gradual at 5 A/cm$^2$ in **Fig. 5e**. These different distributions imply very different distributions of the oxygen potential across the YSZ electrolyte.

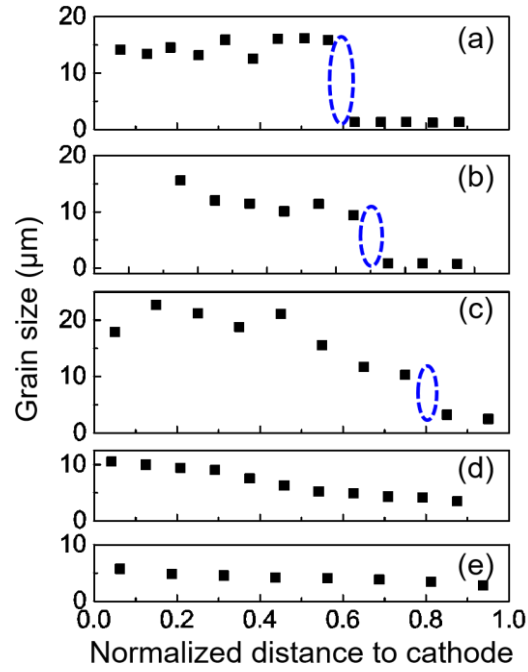

**Figure 6** Grain size distribution of electrically tested YSZ. All are 8YSZ with Pt-paste electrodes, unless noted otherwise. (a) 1.5 mm sample under 50 A/cm$^2$ for 24 h at 1250 °C in air, (b) 0.9 mm 3YSZ under 25 A/cm$^2$ for 20 h at 1200 °C in air, (c) 1.1 mm under 9.7 A/cm$^2$ for 8 h at 1200 °C in air with Pt-wire electrodes (see **Fig. 7**), (d) 0.9 mm under 10 A/cm$^2$ for 4 h at 1300 °C in 5% H$_2$, and (e) 1.3 mm under 5 A/cm$^2$ for 4 h at 1300 °C in 5% H$_2$. Blue oval marks location of discontinuous change in grain size in (a-c).

(2) Raising $\mu''_{O_2} - \mu'_{O_2}$ using bad electrode kinetics

A biased microstructure across the sample developed under a current density of about 10 A/cm$^2$ in air when Pt wires were directly used as electrodes, without any Pt pastes. Very large grains on the cathode side are easily visible under an optical microscope as shown in **Fig. 7a**, which also contains many black dots/patches—they are cracks and pits where grains had become separated during polishing. Under an SEM, we determined a grain size about 2.5 μm (**Fig. 7b**) on the anode side and 20 μm (**Fig. 7c,** which also show many missing grains) on the cathode side. A grain size transition is apparent in **Fig. 7a** at about 200 μm from the anode, bringing the grain size from ~3 μm (similar to the original one) to ~10 μm as shown by the grain size distribution in **Fig. 6c**. This relatively abrupt microstructure transition signals a correspondingly abrupt transition in the oxygen potential. Importantly, although these tests were conducted in air just like in the previous tests reported in Ref. 6 and 7, the grain-growth enhancement occurred at a smaller current density ~10 A/cm$^2$ than at >20 A/cm$^2$ in the previous tests. Indeed, in the previous studies, a microstructure comparable to **Fig. 7** (e.g. **Fig. 4** and **6** of Ref. 7) did not develop until ~50-60 A/cm$^2$. Since the only difference between the two studies is the electrodes, we may conclude that bad electrodes can trigger cathodically enhanced grain growth much easier. Once again, such cathodic enhancement in air signals that a large over potential has developed at the cathode of a YSZ cell with bad electrodes, corresponding to case (iv) in **Fig. 1b**. Meanwhile, the absence of grain-growth enhancement on the anodic side is also consistent with case (iv) in **Fig. 1b**. In comparison, our previous air studies in Ref. 6 and 7 using good electrodes correspond to case (ii) especially when a small current density (<20 A/cm$^2$) is used.

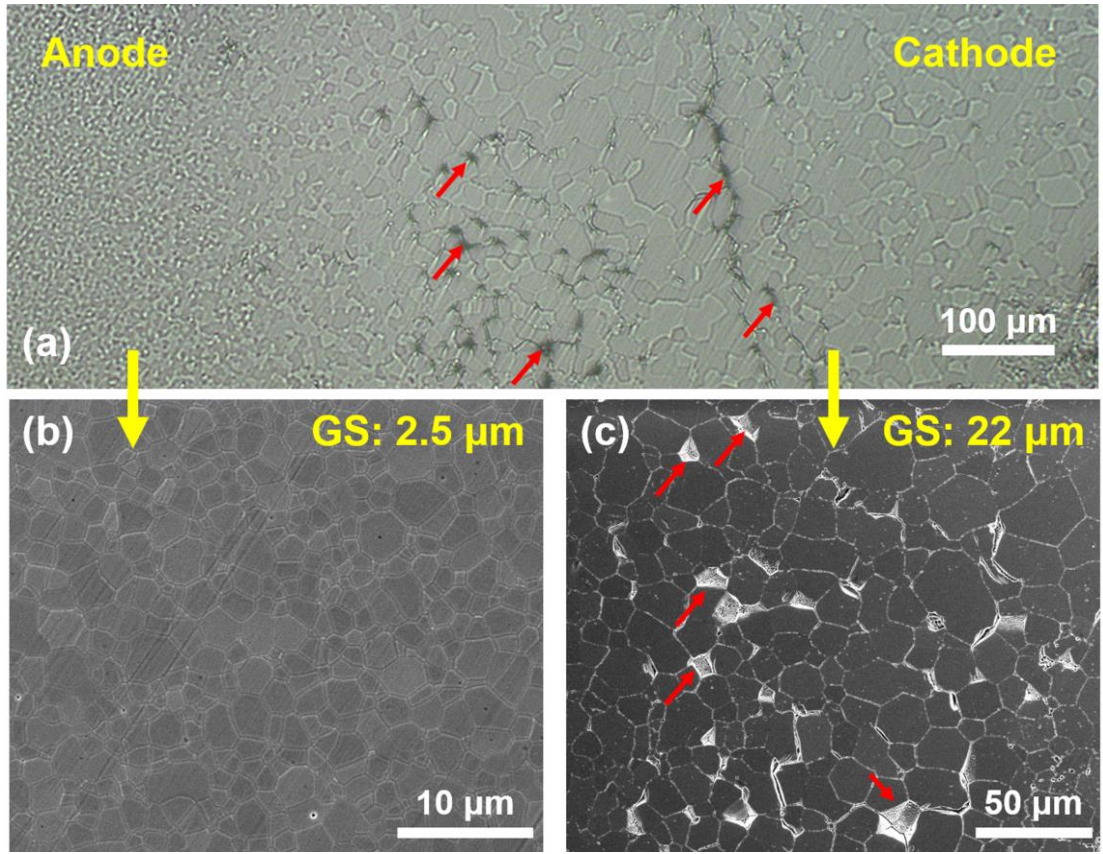

**Figure 7** (a) Optical microscopy micrograph of 8YSZ sample electrically loaded in air under 6.0 V for 8 h. The current density varied from 5.0 to 10.5 A/cm$^2$ with an average value of 9.7 A/cm$^2$. Furnace temperature: 1210 °C; sample thickness: 1.1 mm. Also shown are representative high-magnification SEM images of (b) anode and (c) cathode side. The measured grain sizes are marked on the upper right corners. Red arrows in (a) and (c) mark selected regions with cracks and pits of missing grains that fell off during polishing.

(3) Pore/void formation as indicator of internal reactions

In **Fig. 4-5**, there are numerous pores/voids on the grain boundaries of electrically tested samples on both cathode and anode sides. When tested in air, such extensive cavitation can only be found on the cathode side but not on the anode side.

This was already evident in **Fig. 8a** of **Paper I**[6] where pores and pits formed in air-tested 3YSZ on the cathode side but not on the anode side (**Fig. 8d** therein). Here, **Fig. 7a** and **c** for the air-tested poorly-electroded sample provide another example, showing a cathode side with many pits due to poorly adhered grains that fell apart during polishing, compared to an anode side (**Fig. 7b**) that has neither pores nor pits. A closer view of the pits in 8YSZ in **Fig. 4f** and **Fig. 7c**, and 3YSZ of **Paper I** (**Fig. 8** therein)[6] is presented in **Fig. 8b-d**, which clearly reveals that poor cohesion is the result of extensive grain boundary cavitation, forming highly developed and coalesced pores. Also shown in **Fig. 8a** is an enlarged view of the same region on the cathode side of **Fig. 4f** where grain de-cohesion occurs: Even though most grain boundaries remain intact, they are already heavily decorated by many pores. Finally, we make two notes. (i) **Fig. 8** demonstrates that cavitation mostly occurs in grains that have grown away from the initial size, and its extent is strongly correlated to enhanced grain growth, hence a negative over potential. (ii) Pt anode can easily detach from the YSZ electrolyte after the electrical testing, indicating a poor adhesion or possible cavitations at the anode/YSZ interface.

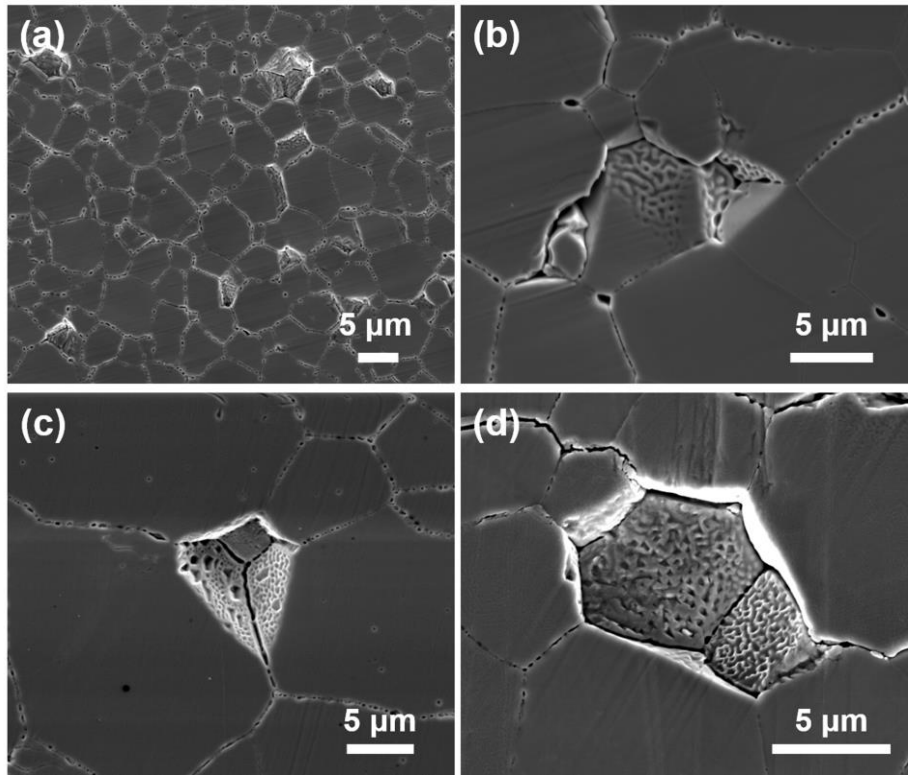

**Figure 8** (a) Pore-decorated grain boundaries developed in same 8YSZ sample shown in **Fig. 4f**, 4.3 V for 4 h in 5% $H_2$ under 10 $A/cm^2$. Furnace temperature: 1300 °C; sample thickness: 0.9 mm. (b) Same sample as (a) at higher magnification showing intergranular cavitation leading to grain fallout and pitting during polishing. (c) Same 8YSZ sample as **Fig. 7c** with cavitation and grain fallout. 6.0 V for 8 h in air, with average current density of 9.7 $A/cm^2$. Furnace temperature: 1200 °C; sample thickness: 1.1 mm. (d) 3YSZ sample showing cavitation and grain fallout. Electrically loaded under 25 $A/cm^2$ at 6.5 V for 20 h. Furnace temperature: 1200°C; sample thickness: 0.9 mm.

## IV. Discussion

Our results using grain size as a marker of local polarization (of oxygen potential) verified all four cases (i-iv) in **Fig. 1b** with only one exception: Anode polarization in

(iii) is apparently negative, since enhanced grain growth is observed therein. Another interesting observation is widespread grain boundary cavitation at both the cathode and the anode side, e.g., in 5% $H_2$ at 1-10 A/cm$^2$ (**Fig. 4**). This is surprising because cavitation has been commonly assumed to result from pressurized oxygen bubbles[11,15-20], thus it should be limited to the anode side and never happen on the cathode side, certainly not in reducing atmospheres. To resolve these contradictions, we will first recap our results on polarization in the broader context of electrolyte, interface, and zirconia cells, then discuss the possible origin of negative polarization at the anode. We will next reconcile the finding of reducing-atmosphere cavitation and discuss other aspects of microstructure and electrochemistry in YSZ ceramics.

(1) Electrode, polarization and kinetics

As mentioned in the **Introduction**, polarization arises from interface control. Hebb provided the following classic examples[21]: (a) When both electrodes are made of Ag for a mixed-conducting electrolyte $Ag_2S$, there is mixed conduction of $Ag^+$ and electron; (b) when Ag and $Ag_2S$ are sandwiched by AgI, which is electron-blocking, there is only $Ag^+$ conduction; (c) when both electrodes are made of Pt, which is $Ag^+$-blocking, there is only electron conduction. Therefore, how a mixed-conducting electrolyte conducts depends on the electrode characteristics.

Usually, 8YSZ is an oxygen ion conductor with little electronic conduction[10,15]. As such it is used as an electrolyte in solid oxide fuel/electrolysis cells (SOFC/SOEC), where the emphasis is on finding ideal electrodes (e.g., with good catalysts, porous electrodes, long triple-phase boundaries for gas/ion/electron exchange) to enable fast interfacial reactions even at a large current/power, thus minimizing voltage loss due to over potential. However, a reduced or over-oxidized YSZ is a mixed conductor. One example is current blackening, e.g., when forcing a current through a YSZ in argon,[22,23] but as demonstrated here it can also be achieved in air by employing a very large current or using poor electrodes. Such blackened YSZ must have developed polarization inside. Another example is a glass-sealed YSZ, which cannot exchange oxygen with the environment[23], so there are only electrons to conduct. At the steady state, $O^{2-}$ must accumulate at the blocked anode to build a positive oxygen potential therein. Therefore, once polarization develops, YSZ behaves like a mixed conductor.

One interesting correlation common to **Fig. 1** and to all the polarized YSZ is: the more oxygen blocking the electrodes, the more significant the electron conduction, the more severe the polarization, and the heavier the electrolyte reduction—at the cathode. This correlation can explain the various grain structures observed by noting the strong dependence of grain growth on severe YSZ reduction. For example, to obtain ion-blocking electrodes, we have used Pt wires, which provide few triple-phase boundaries and cover only a small portion of the electrolyte surfaces. (A glass seal, which will wet the grain boundary and change the grain growth kinetics, was not suitable.) To deprive the electrodes of oxygen supply, we have also used an oxygen-lean environment, which again forces polarization to develop. Since these conditions are opposite to the ones desired for an efficient YSZ SOFC/SOEC, in a sense we are studying degraded zirconia cells. It is interesting to note that the electrode kinetics is maximized in an efficient SOFC/SOEC but its electrolyte-cation kinetics is minimized, resulting in a stable microstructure. In contrast, in our degraded cell, the electrode kinetics is minimized but the electrolyte-cation kinetics is maximized, resulting in rapid grain growth.

But how is enhanced grain growth possible at the anode in the case of electrical testing under Ar or 5% $H_2$? As mentioned in the **Introduction**, due to interface control, any drop in the oxygen potential $\mu_{O_2}$ relative to the potential on the other side of electrode should always be on the downstream side of the oxygen flow, which is from the cathode to the anode in **Fig. 1b**. However, the oxygen potential profile need not be linear as depicted in **Fig. 1b**. As will be explained later, the profile in the case of severe polarization is relatively flat on the cathode side but very steep at close to the anode end. This is illustrated by the numerical results shown in **Fig. 9** for case (iii-iv). (More discussion on this later in the section.) Since the anode regions in **Fig. 4-5** are typically taken at ~50-100 μm away from the electrode, they do experience a substantial negative potential according to **Fig. 9**, which is consistent with enhanced grain growth.

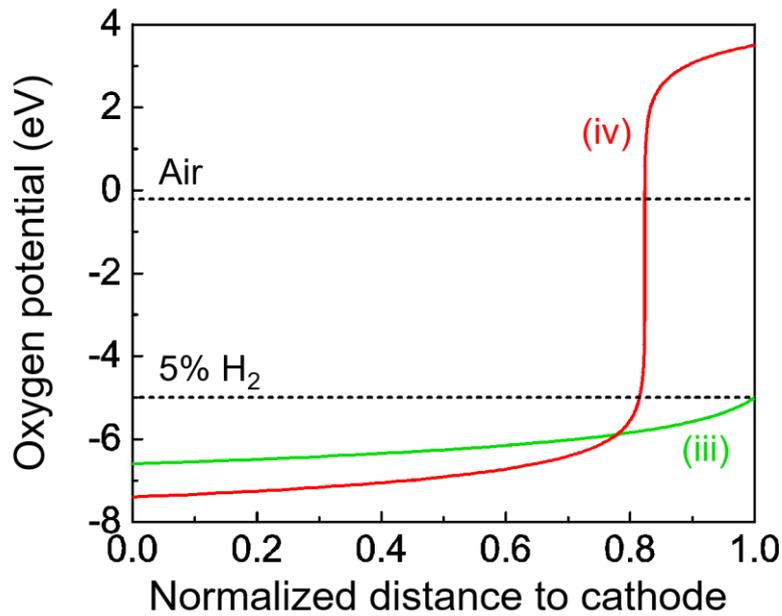

**Figure 9** Calculated oxygen potential profiles inside 8YSZ electrolyte as a function of normalized distance to cathode. Red curve: 1.1 mm thick, under 9.7 A/cm² in air, sample temperature of 1300 °C, $\mu_{O_2}$ from −7.4 eV to 3.5 eV, corresponding to case (iv) in **Fig. 1b**. Green curve: 1.3 mm thick, under 5 A/cm² in 5% H$_2$, sample temperature of 1300 °C, $\mu_{O_2}$ from −6.6 eV to 5.0 eV, corresponding to case (iii) in **Fig. 1b**.

(2) Pore/void nucleation and growth

Grain boundary cavitation has been reported in the literature of SOEC and SOFC[17-20]. Cavitation is usually seen on the anode side of SOEC, especially when the anode is exposed to air under a large current, which causes a very positive oxygen potential. However, during extended testing, widespread cavitation is also seen all the way to the cathode side[19]. Another common observation is anode delamination, which we also observed in our experiments. But cavitation leading to delamination may occur inside the electrode and not in the electrolyte, thus it is not relevant to this study in which we only studied the microstructure inside the electrolyte, at some distance away from the electrodes. These cavities have been thought to be pressurized oxygen bubbles,[11,15-19] which most readily form next to the anode, although they may also appear elsewhere given enough time for bubble nucleation and growth.

What makes our observations of cavitation very different is that it predominantly occurred under a negative oxygen potential. It is also widespread: Referring to **Fig. 4-5**, we can verify cavitation in all cathode regions, and the only anode region that does not have substantial cavitation is at 1 A/cm$^2$ in Ar testing, in **Fig. 5a.** Referring to **Fig. 3**, we can see that where cavitation occurs, there is almost always enhanced grain growth, which signals a negative oxygen potential. Comparing **Fig. 5a** (without cavitation) and **Fig. 4a** (with cavitation), we can also identify the critical level of reduction required for cavitation: It is between testing in 5% H$_2$ (**Fig. 4a**) and in Ar (**Fig. 5a**), at 1 A/cm$^2$, at the anode side. Lastly, we further found cavitation on the cathode side of air-tested samples (**Fig. 7c**), again accompanied by enhanced grain growth indicative of a negative oxygen potential because of poor electrode (case (iv) in **Fig. 1b** and **Fig. 9**). To summarize, cavitation is characteristically accompanied by enhanced grain growth and a negative oxygen potential, thus it cannot involve pressurized oxygen bubbles.

We propose these cavities form by the condensation of supersaturated oxygen vacancies. This is accompanied by reprogramming of lattice cations, pushed to the cavity surfaces. Very likely, these cations are reduced to metallic Zr/Y and absorbed on the cavity surfaces, although metallic Zr/Y is difficult to verify since subsequent cooling and exposure to air can easily re-oxidize it. The above reaction can be written as

$$\text{Zr}_{\text{Zr}}^{\times} + 2\text{V}_{\text{O}}^{\cdot\cdot} + 4e = \text{Zr}_{\text{absoprted}} \quad (1a)$$

or, equivalently,

$$\text{Zr}_{\text{Zr}}^{\times} + 2\text{O}_{\text{O}}^{\times} = \text{Zr}_{\text{absoprted}} + \text{O}_2 \quad (1b)$$

Clearly, its driving force $\Delta g$ should be $k_\text{B}T\ln P\text{O}_2$, with $k_\text{B}$ being the Boltzmann constant and $T$ the absolute temperature, if $\text{Zr}_{\text{absorbed}}$ is in the standard state. This, of course, is not, because the energy of $\text{Zr}_{\text{absorbed}}$ will increase by at least about 1 eV due to underbonding. (To obtain the estimate, we assume a surface energy $\gamma$ of 1 J/m², and from the surface Zr density of $\text{ZrO}_2$, $8\times10^{18}$ Zr/m², we can at least attribute the surface energy to 0.78 eV per surface Zr.)

For realistic nucleation of a vacancy void, its activation energy

$$\Delta G = C \frac{16\pi\gamma^3}{3(\Delta g)^2} \quad (2)$$

should not much exceed $40k_\text{B}T$ where $C=1$ in homogeneous nucleation and $1/C >1$ is a potency factor that characterizes a nucleation site in heterogeneous nucleation. If we let $C=1$ and $T=1{,}500$K, we estimate $\Delta g = 0.9$ eV, which is quite accessible in our samples given the large negative oxygen potential of the order of several eV (**Fig. 9**). Therefore, we expect ready void nucleation in most cases in our samples, except where oxygen's negative potential is very low, e.g., near the anode in Ar-tested samples at 1 A/cm². The critical oxygen potential for nucleation is about −3 to −4 eV at 1,300 °C (see **Fig. 9**).

We next address void growth, which is an oxygen (vacancy) diffusion process following a growth law $r^2 \sim D_O \Delta g$.[24] Here, a diffusion distance of the order of $r$ is assumed for a spherical void, and $D_O$ is the diffusivity of oxygen vacancy, which is fast, probably similar in lattice and along the grain boundary, and insensitive to $\Delta g$. Therefore, the growth dependence on $\Delta g$ is weak, and the growth disparity in different regions of different $\Delta g$ should not be very pronounced. This is consistent with the insets of **Fig. 4b-f** and **Fig. 5c-f**, which show relatively similar void geometry and population in different regions and under rather different electrical loading conditions (hence $\Delta g$). Such insensitivity to oxygen potential is in sharp contrast with the extreme sensitivity in grain growth kinetics. This is because grain growth is controlled by cation diffusion, which is hugely sensitive to the degree of reduction, unlike $D_O$. Therefore, we conclude that in our experiments void formation on the anode side is under nucleation control, and elsewhere under growth control. This scenario is schematically depicted in **Fig. 10a** for electrical loading in a strongly reducing atmosphere.

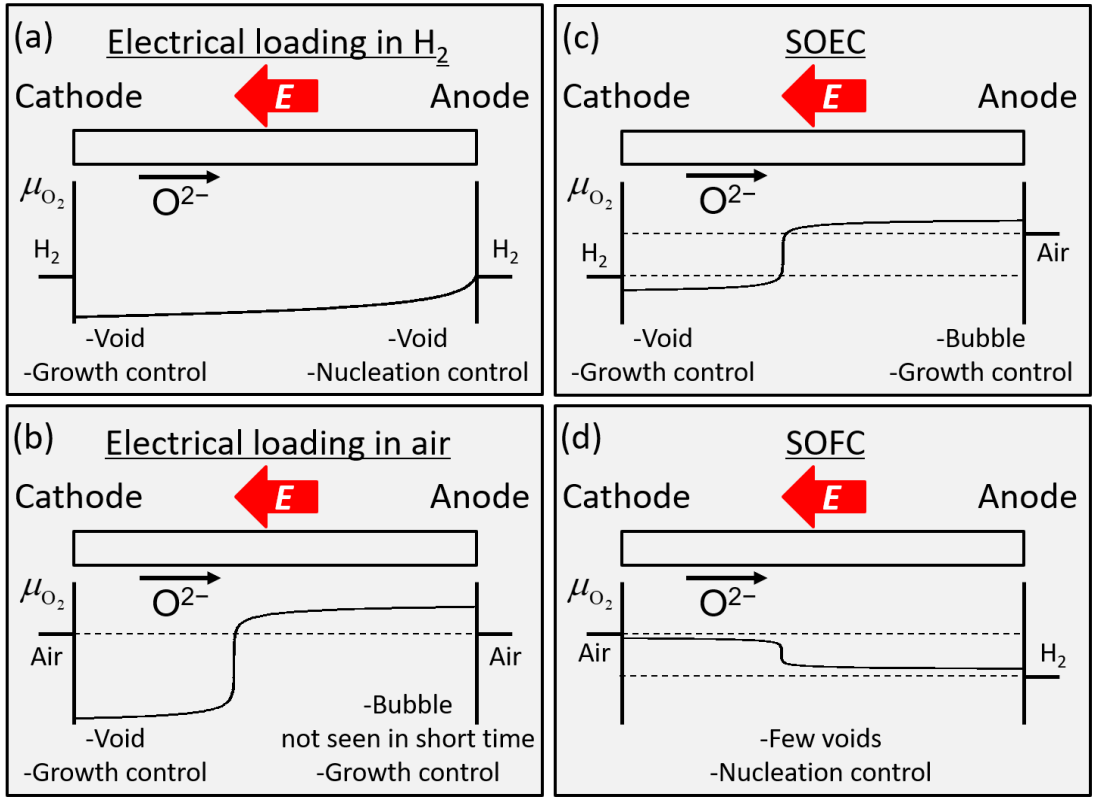

**Figure 10** Schematics of oxygen potential distribution and cavitation for electrically loaded YSZ cell in (a) H$_2$, (b) air, (c) SOEC, and (d) SOFC.

When tested in air with a large current density or a set of poor electrodes, a large negative oxygen potential also develops on the cathode side, but the anode side should have a positive oxygen potential. Therefore, while vacancy condensation can again form voids on the cathode side, any cavitation on the anode side must come from pressurized oxygen by the following reaction of evaporating lattice or interstitial O$^{2-}$ and releasing their electrons to a less-pressurized environment,

$$2O_O^\times = O_2 + 2V_O^{\bullet\bullet} + 4e \qquad (3a)$$

or

$$2O_i'' = O_2 + 4e \qquad (3b)$$

The bubble growth is pressure driven with outward diffusion of $Zr_{Zr}^{\times}$ and $O_O^{\times}$, thus it is controlled by cation diffusion, which is very sluggish in the oxidizing atmosphere where the bubble forms. (For the same number of oxygen anion/vacancy involved, a bubble is bigger than a void because of internal pressure if there is no kinetic limitation.) Therefore, over a short experimental time that does not allow much bubble growth on the anode side, it is possible not to observe cavitation on the anode side in **Fig. 7**. This scenario is schematically depicted in **Fig. 10b**.

Compared to our experiments, a lower current density is encountered in a practical SOEC, whose anode is exposed to an oxidizing atmosphere and whose cathode is exposed to a reducing atmosphere as shown in **Fig. 10c**. Its oxygen over potential should be similar to that in **Fig. 10b** but is less pronounced, so the tendency for cavitation should also be similar but with a slower nucleation and growth rate. However, the electrode kinetics of SOEC may suffer from the relatively low operation temperature compared to the one used in our experiments, which can exacerbate polarization. Moreover, the duration of SOEC operation is long, so, over time, oxygen bubbles may become fully developed on the anode side. Meanwhile, vacancy voids will also nucleate after a long time. This is consistent with the report of cavitation in the SOEC literature.[19]

SOFC experiences an even lower current density than SOEC, with an opposite atmospheric bias to that of SOEC. So, there is always a negative overpotential as depicted in **Fig. 10d**. The oxygen potential inside an SOFC is always bounded by the two terminal oxygen potentials, $\mu_{O_2}^{I}$ and $\mu_{O_2}^{II}$, which are usually air on the cathode side and a fuel (e.g., H$_2$) on the anode side.[9,13] Any cavitation is likely to be on the anode side, due to vacancy voids.

(3) Enhanced grain boundary mobility and non-stoichiometry

From cavitation, one can estimate the concentration of supersaturated oxygen vacancies. First, attribute the difference between the volume $V_{ZrO_2}$ of ZrO$_2$ (A unit cell containing 4 Zr and 8 O is 132 Å$^3$) and the volume $V_{Zr}$ of metallic Zr (93 Å$^3$ for 4 Zr with hexagonal packing) to 2 oxygen vacancies. Next, take the volume faction of vacancy voids in the sample to be $3Ar/R$, where $A$ is the area fraction of grain boundary covered by cavities, which we assume is the percentage of cavitated grain boundaries in **Fig. 8**, $R$ is the grain radius taken as one half of the grain size, and $r$ is the radius of a spherical void. The fraction of oxygen sites that become vacant and forms voids to relieve vacancy supersaturation and restore equilibrium is thus $3A\frac{r}{R}\frac{V_{ZrO_2}}{V_{ZrO_2}-V_{Zr}}$. This fraction falls in the range of 0.03-0.3 if we let $A$ be 1, $R$ be 3-10 μm and $r$ be 0.3-1 μm. In 8YSZ, this number may exceed the fraction of vacant oxygen sites created by Y$_2$O$_3$ doping (3.7%), confirming a large vacancy supersaturation indicating significant reduction. In the following, this number is referred to as nonstoichiometry $\delta$.

To quantify the effect of nonstoichiometry, which is the oxygen deficit relative to the fully oxidized doped composition, on grain growth, we first estimate the grain boundary mobility using the parabolic growth law, which has been confirmed many times in fluorite structure oxides[6-9]

$$d^2 - d_0^2 = 2M\gamma t \qquad (4)$$

Here $d$ denotes the grain size, $d_0$ the initial grain size before testing, $M$ the grain boundary mobility, and $\gamma$ the grain boundary energy taken as 0.3 J/m² in the present work. The calculated grain boundary mobilities are plotted in **Fig. 11**, where our earlier results obtained in air (without electric field, solid line), in 5% H$_2$ (without electric field, dashed line)[6], in air under current densities of 50-60 A/cm² (green squares)[7] and 7.3 A/cm² (orange hexagon) using Pt-paste electrodes, were also included. They vary continuously with no indication of a discontinuous transition. In all, it is possible to vary the grain boundary mobility over 3 orders of magnitude by varying the atmospheres, electrodes and polarization without changing the temperature.

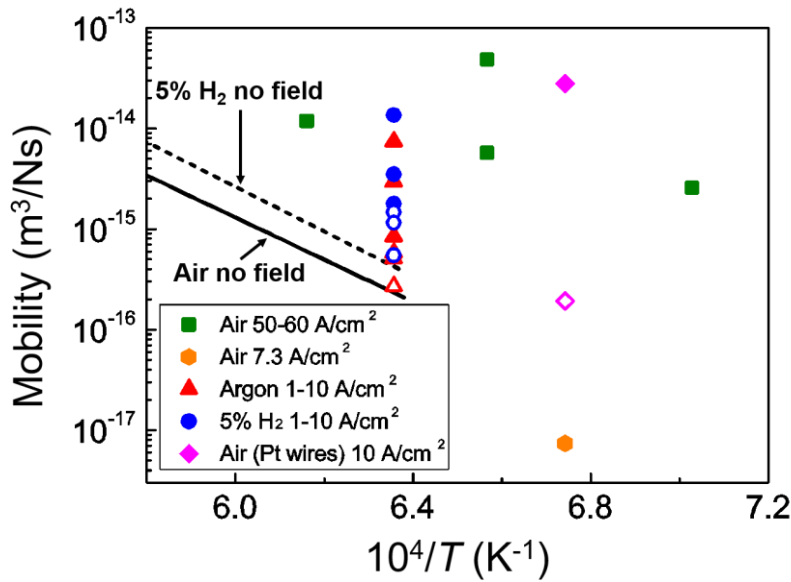

**Figure 11** Arrhenius plot of grain boundary mobility of 8YSZ in air without electric field (solid line), 5% H$_2$ without electric field (dashed line), in air under current densities of 50-60 A/cm² (green squares, for cathode side) and 7.3 A/cm² (orange

hexagon, for both cathode and anode sides) using Pt paste as electrodes, in air under 10 A/cm$^2$ using Pt wires as electrodes (purple diamond, filled symbol for cathode side and open one for anode), in argon under 1-10 A/cm$^2$ using Pt paste as electrodes (red triangles, filled symbols for cathode side and open ones for anode) and in 5% H$_2$ under 1-10 A/cm$^2$ using Pt paste as electrodes (blue circles, filled symbols for cathode side and open ones for anode).

The continuous variation of grain boundary mobility is consistent with a continuous range of non-stoichiometry: There is no reduction-induced phase transition in 8YSZ at high temperatures. To illustrate this, we redraw a figure in **Paper I**[6] in **Fig. 12**, which includes our new data of enhanced mobilities and their expected nonstoichiometry $\delta$ to be consistent with the trend. (A nonstoichiometric 8YSZ is Zr$_{0.852}$Y$_{0.148}$O$_{1.926-\delta}$, a nonstoichiometric 3YSZ is Zr$_{0.942}$Y$_{0.058}$O$_{1.971-\delta}$, a nonstoichiometric undoped ceria is CeO$_{2-\delta}$, and a nonstoichiometric GDC (abbreviation for Gd$_{0.1}$Ce$_{0.9}$O$_{1.95}$) is Gd$_{0.1}$Ce$_{0.9}$O$_{1.95-\delta}$. That is, no reduction when $\delta=0$, vacancies supersaturate when $\delta>0$.) The range of the required $\delta$ to obtain the enhancement factor shown in **Fig. 12** seems reasonable given the estimate of $\delta$ from cavitation. Large nonstoichiometry arises because of the inability of electrode kinetics to match the large electric current demand on the electrodes.

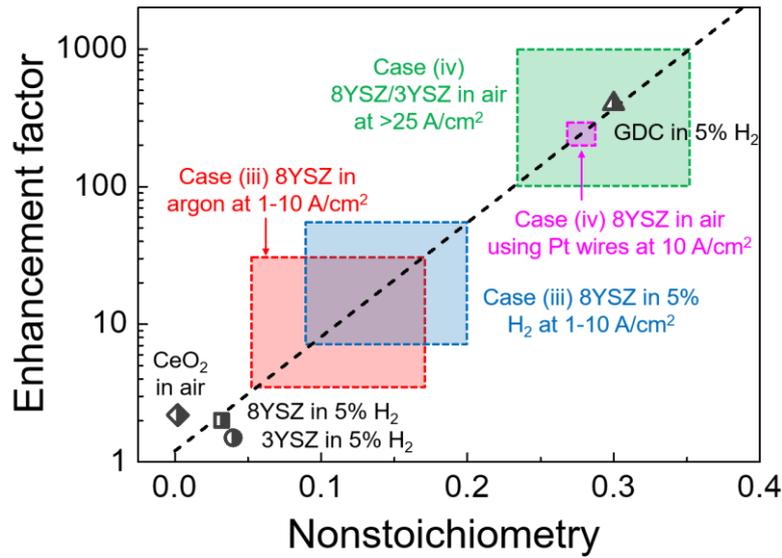

**Figure 12** Correlation between mobility enhancement of grain boundary and nonstoichiometry. The dashed line is to guide the eye. Enhancement factor is defined as the ratio of grain boundary mobility under reducing condition to that under reference oxidizing condition. For 8YSZ, 3YSZ and GDC, the reference oxidizing condition is in air; for $CeO_2$, it is in pure oxygen, all without current.

(4) Oxygen potential transition and internal reactions

To compute the polarization and oxygen potential distribution, one need to consider both defect transport and defect reactions. The latter adds a source/sink term to the transport equation, which may be solved under the assumption of steady state. The prevalence of cavitation in our samples (and in the literature[17-20]) signals internal reactions between ionic defects and electrons/holes regardless of the cavitation mechanism. In forming oxygen bubbles, each condensing lattice/interstitial oxygen is neutralized by shedding two electrons. In forming vacancy voids, each condensing oxygen vacancy is neutralized by acquiring two electrons. Electron spin resonance measurements have also found partially ionized oxygen vacancies in reduced YSZ as color centers,[25] suggesting conversion between $V_O^{\cdot\cdot}$, $V_O^{\cdot}$ and $V_O^{x}$, again involving internal reactions between ionic defects and electrons.

The influence of conversion reactions on oxygen potential calculations can be best appreciated by reviewing the case without such reactions. Referring to the **Appendix** and denoting $t_i = j_i / j$ and $t_e = j_e / j$, where $j$ is the total current density, $t_i$ (and $j_i$) and $t_e$ (and $j_e$) are the transference number (current density) for ionic and electronic current, respectively, we rewrite Eq. (A7) as

$$j = \frac{1}{4e}\left(\frac{t_i}{\sigma_i} + \frac{t_e}{\sigma_e}\right)^{-1} \frac{d\mu_{O_2}}{dx} = M \frac{d\mu_{O_2}}{dx} \qquad (5)$$

where $\dfrac{d\mu_{O_2}}{dx}$ is the oxygen-potential gradient between the electrodes, and $M$ is an effective electromobility that accounts for all the carriers and their respective contributions to the total current density. (Electronic current includes both electron current and hole current.) One can integrate Eq. (5) to obtain

$$\mu_{O_2}'' - \mu_{O_2}' = \int_0^L \frac{j}{M} dx \qquad (6)$$

where $\mu_{O_2}'$ and $\mu_{O_2}''$ are the oxygen potentials at the two electrode/electrolyte interfaces, and $L$ is the thickness of the electrolyte. This integral and $t_i + t_e = 1$ uniquely determine the transference numbers. The solution of $\mu_{O_2}$ is next obtained by integrating to any intermediate $x$ between 0 and $L$. This solution has the following feature: Under the same $\mu_{O_2}'$ and $\mu_{O_2}''$, an increasingly larger $j$ can only be achieved by an increasingly larger $M$, which is possible only if the electrolyte is increasingly reduced on the cathode side and increasingly oxidized on the anode side. This forces the oxygen potential distribution to take essentially the two boundary values, $\mu'_{O_2}$ and $\mu''_{O_2}$, over most of the length *L*, except in a narrow transition zone between the two boundary values. (In **Paper I**, we refer to this as a n-i-p configuration.) So the potential distribution is highly non-linear unlike depicted in **Fig. 1b**. Moreover, if the sample thickness is sufficiently large, the highest mobility region (in our experiment, it is the n-type region) must span most of *L*, so the transition is pushed to very near the anode.

It turns out that the effect of internal reactions converting ionic species via addition/association of electronic defects is like that of a chemical buffer: It smooths the oxygen-potential transition between electrodes. It also moves the transition away from the electrode toward the center of the electrolyte. This is consistent with the grain size transition in **Fig. 6**, where the transition zone is not very near the electrode. Numerical calculations considering internal reactions to form species that are not fully ionized are shown in **Fig. 9**.

The details of such calculations will be published elsewhere, but below we shall describe how the boundary oxygen potentials were selected. The two cases studied in **Fig. 9** are (a) a 1.1 mm thick sample under 9.7 A/cm$^2$ in air with a sample temperature of 1300 ºC (the red curve) and (b) a 1.3 mm thick sample under 5 A/cm$^2$ in 5% H$_2$ with a sample temperature of 1300 ºC (the green curve). On the cathode side, the observed grain growth enhancement is about 300× in (a) and 10× in (b), corresponding to $\delta$ of 0.3 and 0.1, respectively, according to **Fig. 12**. Using the

relationship between $\delta$ and $PO_2$ given below for 8YSZ[26] (see **Paper I**[6])

$$\delta^2(\delta+0.074)PO_2^{0.5} = 0.2276\exp(-3.98 \text{ eV}/k_\text{B}T) \qquad (7)$$

we obtain $\mu_{O_2} = -7.4$ eV for $\delta=0.3$ and $\mu_{O_2} = -6.6$ eV for $\delta=0.1$ at 1300 ºC. On the anode side, we expect the oxygen potential in (b) to be very close to that of 5% H$_2$, or $PO_2 \approx 10^{-16}$ atm according to Ref. 27, i.e., $\mu_{O_2} = -RT\ln(PO_2/1\text{ atm}) = -5.0$ eV. In (a), we have no direct information of the oxygen potential $\mu_{O_2}''$ on the anode side, so we chose $\mu_{O_2}''$ so that the grain size transition coincides with the experimental observed location. The numerical solutions in **Fig. 9** illustrate a large negative potential on the cathode side that increases from (a) to (b), and in (b) a very considerably negative potential (−6.0 eV) at ~100 μm (less than 10% of the total thickness) away from the anode. These results are consistent with our observations including grain growth on the anode side when electrically tested in a reducing atmosphere.

(5) Flash sintering in air and graded microstructure

Enhanced cathodic grain growth has been observed in flash sintered 3YSZ.[28] Very large grains up to 50 μm were found next to the cathode under a nominal current density of 10 A/cm² lasting for 60 s at a furnace temperature of 900 ºC (sample temperature much higher), while grains elsewhere remain very small. (We estimated current density using the geometry of the green body, which has a packing density ~50%; if we estimate it using the geometry of the sintered body, which has full density, then the current density is ~17 A/cm²). This likely implies poor electrode kinetics and a large negative potential at the cathode in flash sintering, in air, similar to case (iv) in **Fig. 1b** and **Fig. 7**. Likewise, this may have occurred in other ceramics under flash sintering[29-33] where high voltages, large current densities and non-optimized electrodes (sometimes, just a Pt wire hung over sample's "necks") are common. In comparison, spark plasma sintering is less prone to over potential because (i) it typically uses a lower voltage and (ii) most current passes through the highly conductive graphite dies instead of the ceramic sample.[34]

Our ceramics developed a graded microstructure when loaded by Pt-wire electrodes with a moderate current, in air. Graded microstructures with a grain-size variation between the two electrodes have also been reported after electric loading, some in flash sintering, in monoclinic[35], tetragonal[6] and cubic[7] zirconia, ZnO[31], $MgAl_2O_4$[36], $SrTiO_3$[37] and $UO_2$[38]. Since such microstructure is asymmetric and directional, it cannot originate from a uniform field such as electric field or current density. The reason is: Although electrical potential can direct a directional charge flux, such as current, it does not directly drive non-directional transport of atoms/ions, as in grain growth. (For example, mass movement in a growing grain is radially inward, not uni-directional.) Therefore, grain growth of isotropic materials cannot be directly caused by an electric field. Instead, any modification of grain growth kinetics must come via the polarization of the chemical potential of the rate controlling species, e.g., cations in zirconia[39,40], because of inadequate electrode kinetics. As will become clear in the **Appendix**, the chemical potential of cation is fully determined once the oxygen potential is known. (Here, we ignore the distinction between Zr and Y, which is a good approximation in cubic YSZ that forms a solid solution over a wide range of Zr/Y composition.) Therefore, it is the local oxygen potential that determines the microstructure, or conversely, a graded microstructure is a reflection of a graded oxygen potential. This is the same premise of the present work: The grain size is a marker of polarization.

What is the state variable to use to control the development of a graded microstructure in electrical loading? Clearly, neither voltage nor current density alone is the state variable, since the outcome is strongly dependent on the atmosphere of the environment. The combined specification of the electrical loading condition (voltage and current) and environmental atmosphere is not the state variable either, since the outcome depends on the type of electrodes used. Furthermore, it is not sufficient to just specify the boundary conditions of the thermodynamic potential $\mu_{O_2}$, since the steady-state solution also depends on the current density (see Eq. (6)). On the other hand, once the boundary conditions of $\mu_{O_2}$ at the two electrode-electrolyte interfaces are specified, then there is a unique solution of $\mu_{O_2}$ for a given steady state current density. Thus, it is $\mu_{O_2}$ that is the state variable in the transport equation, and $j$ is like a material constant (similar to permittivity in the electrostatic potential problem) that defines the nature of the transport equation. Varying the atmospheres and the electrode kinetics/materials/geometry can drastically alter the boundary conditions even though the sample is identically loaded electrically. Keeping the boundary condition of $\mu_{O_2}$ the same but varying the current can also drastically alter the $\mu_{O_2}$ distribution. Both will result in different graded microstructures.

Lastly, graded structure produced by passing a large current is likely to be accompanied by grain boundary cavitation or incipient cavitation. If electrical loading is performed in air, then the scenario depicted in **Fig. 10b** will hold and degradation of mechanical properties including intergranular fractures[41] will result. This may be an issue for ceramics sintered under a large current density.

## V. Conclusions

(1) Grain size can be used as a microstructural marker to detect electrode polarization in YSZ. Electrode polarization can severely reduce YSZ and enhance grain growth by 1,000 times, both in reducing atmospheres and under a large current in air.

(2) Extensive cavitation due to oxygen-vacancy condensation on grain boundaries can cause loss of grain-to-grain adhesion in electrically reduced YSZ. It also signals internal chemical reactions between ionic defects and electrons, which can buffer the development of oxygen potential gradient inside a highly polarized electrolyte.

(3) To describe a steady state of a mixed conductor, both the current density and the boundary values of the thermodynamic potentials need to be specified. They both sensitively influence the solution of the field equation, which is highly non-linear because conductivities are strongly dependent on oxygen potential.

**Acknowledgements**

This work was supported by the Department of Energy (BES grant no. DEFG02-11ER46814) and used the facilities (LRSM) supported by the U.S. National

Science Foundation (grant no. DMR-1120901).

**Appendix**

Consider a YSZ cell shown in **Fig. A1a** with the *x*-direction pointing to the right, having negative ionic and electronic current flowing from the anode to the cathode.[11] (The $O^{2-}$ flow and electron flow are to the right.) The current density is driven by the respective electrochemical potential, $\tilde{\mu}_{O^{2-}}$ for $O^{2-}$ and $\tilde{\mu}_e$ for electron,

$$j_e = \text{electronic current density} = \frac{\sigma_e}{e} \frac{d\tilde{\mu}_e}{dx} \qquad (A1)$$

$$j_i = \text{ionic current density} = \frac{\sigma_i}{2e} \frac{d\tilde{\mu}_{O^{2-}}}{dx} \qquad (A2)$$

where $\sigma_e$ denotes electronic conductivity, which could come from both electrons and holes, $\sigma_i$ denotes ionic conductivity, and $e$ is the elementary charge. To proceed, we first represent $\tilde{\mu}_{O^{2-}}$ in terms of oxygen potential $\mu_{O_2}$ and $\tilde{\mu}_e$. This is aided by recognizing the chemical reaction,

$$O^{2-} = \tfrac{1}{2} O_2 + 2e \qquad (A3)$$

which is assumed to attain local equilibrium so that

$$\mu_{O^{2-}} = \tfrac{1}{2} \mu_{O_2} + 2\mu_e \qquad (A4)$$

Here, $\mu_{O^{2-}}$ denotes the chemical potential of $O^{2-}$, and the chemical potential of electron $\mu_e$ is related to $\tilde{\mu}_e$ by

$$\tilde{\mu}_e = \mu_e - e\phi \qquad (A5)$$

with $\phi$ being the electrostatic potential. Therefore, we obtain

$$\tilde{\mu}_{O^{2-}} = \tfrac{1}{2} \mu_{O_2} + 2\tilde{\mu}_e \qquad (A6)$$

The ionic current density is thus

$$j_i = \frac{\sigma_i}{4e} \frac{d\mu_{O_2}}{dx} + \frac{\sigma_i}{e} \frac{d\tilde{\mu}_e}{dx} = \frac{\sigma_i}{4e} \frac{d\mu_{O_2}}{dx} + \frac{\sigma_i}{\sigma_e} j_e \qquad (A7)$$

in which the second equality came from Eq. (A1). The different forms for electron current and ionic current, which are parallel paths in an equivalent circuit shown in **Fig. A1b**, are sometimes interpreted as caused by an electromotive force or emf due to the change of the oxygen potential across the circuit. Further denoting $t_i = j_i / j$ and $t_e = j_e / j$, we can rearrange Eq. (A7) to Eq. (5) in the main text.

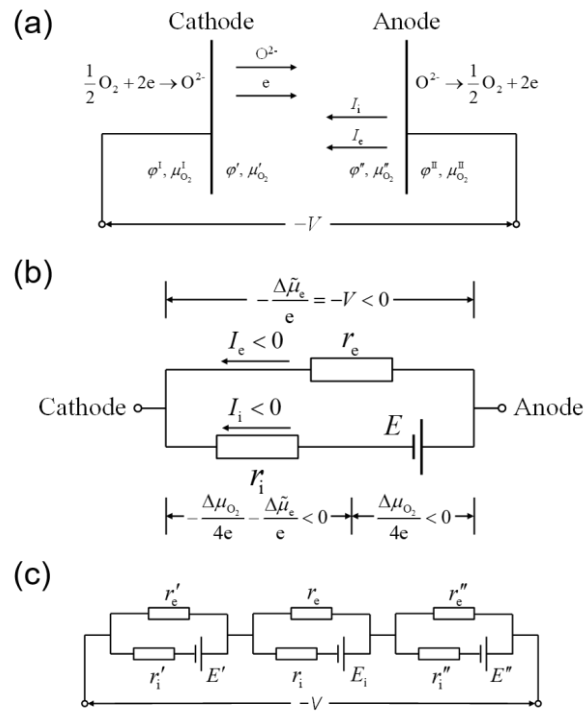

**Figure A1** (a) Schematics for a YSZ cell under an applied voltage $V$, (b) its equivalent circuit. (c) It can be further interpreted as three individual ones for cathode/electrolyte interface, electrolyte, and anode/electrolyte interface.

The above results can be generalized to a cell with three components: electrolyte, cathode/electrolyte interface, and anode/electrolyte interface. To each of them one can assign a set of equivalent ionic and electronic resistances ($r$) that restrict the respective current ($I$), and to each interface a set of electrochemical potentials.

Referring to **Fig. A1a** and **A1c**, which is an equivalent circuit, we write the electronic current as follows

$$I'_e = -\frac{1}{e}\frac{\tilde{\mu}_e^{I} - \tilde{\mu}_e'}{r'_e} \quad (A8)$$

$$I_e = -\frac{1}{e}\frac{\tilde{\mu}_e' - \tilde{\mu}_e''}{r_e} \quad (A9)$$

$$I''_e = -\frac{1}{e}\frac{\tilde{\mu}_e'' - \tilde{\mu}_e^{II}}{r''_e} \quad (A10)$$

where the primed current and resistance are those at the cathode, the double-primed ones are those at the anode, and the interface potentials are similarly denoted by referring to the type of the electrodes. Similarly, following the same notation, we can write the various ionic current as

$$I'_i = -\frac{1}{4e}\frac{\mu_{O_2}^{I} - \mu_{O_2}'}{r'_i} - \frac{1}{e}\frac{\tilde{\mu}_e^{I} - \tilde{\mu}_e'}{r'_i} \quad (A11)$$

$$I_i = -\frac{1}{4e}\frac{\mu_{O_2}' - \mu_{O_2}''}{r_i} - \frac{1}{e}\frac{\tilde{\mu}_e' - \tilde{\mu}_e''}{r_i} \quad (A12)$$

$$I''_i = -\frac{1}{4e}\frac{\mu_{O_2}'' - \mu_{O_2}^{II}}{r''_i} - \frac{1}{e}\frac{\tilde{\mu}_e'' - \tilde{\mu}_e^{II}}{r''_i} \quad (A13)$$

Under the steady state, both electronic and ionic current should be constant throughout the cell assembly. Therefore,

$$I'_e = I''_e = I_e = -\frac{1}{e}\frac{\tilde{\mu}_e^{I} - \tilde{\mu}_e^{II}}{r'_e + r_e + r''_e} \quad (A14)$$

$$I'_i = I''_i = I_i = -\frac{1}{4e}\frac{\mu_{O_2}^{I} - \mu_{O_2}^{II}}{r'_i + r_i + r''_i} - \frac{1}{e}\frac{\tilde{\mu}_e^{I} - \tilde{\mu}_e^{II}}{r'_i + r_i + r''_i} \quad (A15)$$

Hence, with the aid of Eq. (A8) and (A14-15), $\mu'_{O_2}$ in Eq. (A11) can be expressed as

$$\mu'_{O_2} = \mu^I_{O_2} - 4e(r'_i|I_i| - r'_e|I_e|)$$

$$= \mu^I_{O_2} - \frac{r'_i(\mu^I_{O_2} - \mu^{II}_{O_2})}{r'_i + r_i + r''_i} - \frac{4r'_i(\tilde{\mu}_e^{\ I} - \tilde{\mu}_e^{\ II})}{r'_i + r_i + r''_i} + \frac{4r'_e(\tilde{\mu}_e^{\ I} - \tilde{\mu}_e^{\ II})}{r'_e + r_e + r''_e} \quad \text{(A16)}$$

$$= \mu^I_{O_2} - \frac{r'_i(\mu^I_{O_2} - \mu^{II}_{O_2})}{r'_i + r_i + r''_i} - 4eV\left(\frac{r'_i}{r'_i + r_i + r''_i} - \frac{r'_e}{r'_e + r_e + r''_e}\right)$$

Likewise, $\mu''_{O_2}$ can be expressed as

$$\mu''_{O_2} = \mu^{II}_{O_2} + 4e(r''_i|I_i| - r''_e|I_e|)$$

$$= \mu^{II}_{O_2} + \frac{r''_i(\mu^I_{O_2} - \mu^{II}_{O_2})}{r'_i + r_i + r''_i} + \frac{4r''_i(\tilde{\mu}_e^{\ I} - \tilde{\mu}_e^{\ II})}{r'_i + r_i + r''_i} - \frac{4r''_e(\tilde{\mu}_e^{\ I} - \tilde{\mu}_e^{\ II})}{r'_e + r_e + r''_e} \quad \text{(A17)}$$

$$= \mu^{II}_{O_2} + \frac{r''_i(\mu^I_{O_2} - \mu^{II}_{O_2})}{r'_i + r_i + r''_i} + 4eV\left(\frac{r''_i}{r'_i + r_i + r''_i} - \frac{r''_e}{r'_e + r_e + r''_e}\right)$$

where $V = \dfrac{\tilde{\mu}_e^{\ I} - \tilde{\mu}_e^{\ II}}{e}$ is the externally applied voltage. In our experiment, an overall negative current is imposed by applying a positive voltage $V$ to anode, and the cell is placed in a uniform atmosphere, where $\mu^I_{O_2} = \mu^{II}_{O_2}$. So the above results are reduced to

$$\mu'_{O_2} = \mu^I_{O_2} - 4eV\left(\frac{r'_i}{r'_i + r_i + r''_i} - \frac{r'_e}{r'_e + r_e + r''_e}\right) \quad \text{(A18)}$$

$$\mu''_{O_2} = \mu^{II}_{O_2} + 4eV\left(\frac{r''_i}{r'_i + r_i + r''_i} - \frac{r''_e}{r'_e + r_e + r''_e}\right) \quad \text{(A19)}$$

in which the ratio of $\dfrac{r'_i}{r'_i + r_i + r''_i}$ (or $\dfrac{r''_i}{r'_i + r_i + r''_i}$) and $\dfrac{r'_e}{r'_e + r_e + r''_e}$ (or $\dfrac{r''_e}{r'_e + r_e + r''_e}$) at the two electrodes and their interfaces become crucial. In general, the electronic resistance ratio in the bracket on the right-hand side is quite small because the electrode is thin and electrodes have low electronic resistivity. So it is the ionic resistance ratio that determines the oxygen potential difference across the electrode/interface.

We now consider the following cases.

**(i) Ideal electrodes**: We define ideal electrode by demanding its oxygen potential the same as that of the atmosphere. To fulfill this, in general, the electrode/interface must have no ionic or electronic resistance. Having no ionic resistance implies not only no resistance to $O^{2-}$ transport but also infinitely fast reactions to incorporate or release oxygen. Therefore, a small current is usually a necessity to realize ideal electrodes. In **Fig. 1b**, it corresponds to the case of a flat oxygen potential by the black dash lines.

**(ii) YSZ cell in air with a small current**: In this case, the electrolyte is a good ionic conductor but poor electronic conductor, and vice versa for the electrodes and their interfaces. In addition, the electrode/electrolyte interfaces are much thinner than the electrolyte. Therefore, $r_e'$ and $r_e''$ should be much smaller than $r_e$. In contrast, $r_i'$ and $r_i''$ can have a finite value that is comparable with $r_i$. Therefore, $\frac{r_e'}{r_e' + r_e + r_e''}$ and $\frac{r_e''}{r_e' + r_e + r_e''}$ are negligibly small but $\frac{r_i'}{r_i' + r_i + r_i''}$ and $\frac{r_i''}{r_i' + r_i + r_i''}$ need to be considered. So Eq. (A16-A17) can be approximately written as

$$\mu_{O_2}' \approx \mu_{O_2}^I - 4e\left(V - \left|\frac{\mu_{O_2}^I - \mu_{O_2}^{II}}{4e}\right|\right)\frac{r_i'}{r_i' + r_i + r_i''} < \mu_{O_2}^I \qquad (A20)$$

$$\mu_{O_2}'' \approx \mu_{O_2}^{II} + 4e\left(V - \left|\frac{\mu_{O_2}^I - \mu_{O_2}^{II}}{4e}\right|\right)\frac{r_i''}{r_i' + r_i + r_i''} > \mu_{O_2}^{II} \qquad (A21)$$

In the above, the $\left|\frac{\mu_{O_2}^I - \mu_{O_2}^{II}}{4e}\right|$ term is the Nernst open-circuit potential which should be smaller than the applied voltage $V$ (or else the current will reverse its direction) and it is retained here for completeness even though it is zero in our experiment. Therefore, there will be a lower oxygen potential developed at the cathode/electrolyte interface and a higher oxygen potential developed at the anode/electrolyte interface. The oxygen-potential difference provides the driving force to allow oxygen reaction at the electrode-air interfaces: incorporation from $O_2$ at the cathode and oxygen release to form $O_2$ at the anode. As Eq. (A20-21) show, they vanish for ideal electrodes, defined as ones with negligibly small $r_i'$ and $r_i''$ compared to $r_i$, giving $\mu_{O_2}'$ and $\mu_{O_2}''$ identical to $\mu_{O_2}^{I}$ and $\mu_{O_2}^{II}$, respectively. This is schematically shown as the blue line in **Fig. 1b**, and it likely corresponds to our experiments in air with a small current density, without obvious grain growth enhancement.

**(iii) YSZ cell in hydrogen:** The same argument of the driving force can be applied here, which dictates a difference in oxygen potential across the two atmosphere-electrode interfaces, but they are shifted downward because the oxygen potential of hydrogen is lower. On the cathode side, because of the scarcity of oxygen in the atmosphere, it is difficult to incorporate oxygen into the cell, so $r_i'$ is likely very high, making the ratio $\dfrac{r_i'}{r_i' + r_i + r_i''}$ dominate over the ratio of $\dfrac{r_e'}{r_e' + r_e + r_e''}$ in Eq. (A18). On the anode side, the very slow ionic current and the low oxygen pressure outside makes the release of oxygen relatively easy, so $r_i''$ is likely very small. This makes the difference in $\mu_{O_2}''$ and $\mu_{O_2}^{II}$ very small. This is schematically shown in **Fig. 1b** by the green curve, and it likely corresponds to our experiment conducted in 5% hydrogen (**Fig. 4**).

**(iv) YSZ cell in air with a large current:** Here, heavy redox of the electrolyte

caused by the large current creates the possibility that substantial electronic conductivity may arise. However, the electrode/electrolyte interfaces are much thinner than the electrolyte. So $r'_e$ and $r''_e$ should still be much smaller than $r_e$. In addition, the large current also implies large $r'_i$ and $r''_i$. So a large potential difference across the interface is expected. This is schematically shown in **Fig. 1b** by the red curve, and it likely corresponds to our experiments in air (**Fig. 7**).